\documentclass[aps,prb,twocolumn,groupedaddress]{revtex4}
\usepackage{graphicx}
\begin{document}

%\preprint{}

\title{Induced decoherence and entanglement by interacting quantum spin baths}


\author{Cheng-Yan Lai$^{1}$}
\author{Jo-Tzu Hung$^{1}$}
\author{Chung-Yu Mou$^{1,2}$}
\author{Pochung Chen$^{1}$}
%\thanks{}
\affiliation{ 1.Department
of Physics, National Tsing Hua University, Hsinchu 30013, Taiwan\\
2. National Center for Theoretical Sciences, P. O. Box 2-131,
Hsinchu, Taiwan}


\date{\today}

\begin{abstract}
The reduced dynamics of a single or two qubits coupled to an
interacting quantum spin bath modeled by a XXZ spin chain is
investigated. By using the method of time-dependent density matrix
renormalization group (t-DMRG), we go beyond the uniform coupling
central spin model and evaluate nonperturbatively the induced
decoherence and entanglement. It is shown that both decoherence and
entanglement strongly depend on the phase of the underlying spin
bath. We show that in general, spin baths can induce entanglement
for an initially disentangled pair of qubits. Furthermore, when the
spin bath is in the ferromagnetic phase, because qubits directly
couple to the order parameter, the reduced dynamics shows
oscillatory type behavior. On the other hand, only for paramagnetic
and antiferromagnetic phases, initially entangled states suffer from
the entanglement sudden death. By calculating concurrence, the
finite disentanglement time is mapped out for all phases in the
phase diagram of the spin bath.
\end{abstract}

%\keywords{}

\maketitle


%%%%%
\section{Introduction}
Spin qubits are promising candidates for quantum information
processing due to their long decoherence and relaxation
time.\cite{PhysRevB.61.12639,ladd:014401} Some schemes, such as
solid state spin qbuits, further enjoy the potential scalability via
the integration with nanotechnology.\cite{PhysRevA.57.120} However,
spin qubits are not totally immune from the ubiquitous decoherence.
To describe the bath that causes the decoherence of spin qubits, it
is known that in some cases, the bath is better modeled by spins
instead of delocalized oscillators, resulting in the so-called spin
baths.\cite{Stamp2004} It has been argued that the influence of spin baths
may be qualitatively different from bosonic baths and non-Markovian
dynamics can easily
emerge.\cite{arXiv:0707.1131,cucchietti:052113,breuer:045323} Due to
the growing interest in spin baths, the decoherence behavior and the
entanglement dynamics of few qubits coupled to spin baths have been
studied extensively in recent years. Early works focus on the
decoherence due to independent spins.\cite{cucchietti:052113} Here
although the proposed model formally resembles a spin boson model,
non-Markovian already emerges even when bath modes are not
interacting.\cite{breuer:045323} In real baths, however, spins are
not independent. It is therefore important to include effects due to
interactions of spins in the bath. Nonetheless, the inclusion of the
intra-spin interaction in the bath complicates the problem and only
for some limited models with high symmetry, exact reduced dynamics
can be identified. \cite{yuan:045331} Beyond models with exact
solutions, approximated dynamics was obtained by using
mean-field\cite{PhysRevA.66.052317} or perturbative
approaches\cite{Yuan2007}to handle more generic models. The most
common model employed in these works is the "central spin model",
where the qubits are uniformly coupled to all spins of the bath.
While analytical derivations are possible in these models, they are
less realistic and are more difficult to be implemented experimentally. A
non-perturbative approach that can capture the non-Markovian effects
induced by interacting spin bath with generic coupling to qubits is
hence highly desirable.

To overcome the difficulty associated with interacting spins, we
utilize the method of time-dependent density matrix renormalization
group (t-DMRG) \cite{vidal:040502,white:076401} to investigate the
reduced dynamics of single or two qubits coupled to an interacting
spin chain. Recently, t-DMRG has been used to study the single qubit
pure dephasing induced by a $XXZ$ anisotropic spin
chain.\cite{rossini:032333} The advantage of t-DMRG is its ability
to calculate reduced dynamics even when the spin bath is not
integrable and the coupling is not uniform. Due to the accumulation
of errors, t-DMRG will eventually run away at large
time\cite{gobert:036102} but this does not impose serious limitation
since for the study of quantum information we are mostly interested
in some smaller time scale. In this work we apply the method of t-DMRG
to investigate both pure dephasing and general decoherence model of
qubits coupled to spin baths. Single qubit decoherence as well as
two qubits (dis)-entanglement dynamics are investigated. It is
shown that both decoherence and entanglement strongly depend on the
phase of the spin bath. In general, we find that spin baths can
induce entanglement for an initially disentangled pair of qubits.
However, when the spin bath is in the ferromagnetic phase, because
qubits directly couple to the order parameter, the reduced dynamics
shows oscillatory type behavior. On the other hand, only for
paramagnetic and antiferromagnetic phases, initially entangled
states suffer from the entanglement sudden death.\cite{yu:140404,Jakobczyk2004}

To quantify the single spin decoherence, we evaluate the evolution
of the Loschmidt echo.\cite{PhysRevLett.91.210403} We analyze the
relation between the short time Loschmidt echo decay parameter and
the quantum phases of the spin bath, as it has been pointed out that
these two are closely related, especially when a symmetry breaking
occurs in the bath.\cite{PhysRevA.66.052317} We use the temporal
evolution of concurrence\cite{Wootters1998} to study the
entanglement dynamics. One important issue of entanglement dynamics
is the possibility of creating entanglement through a common bath
for originally dis-entangled qubits. It has been shown that induced
entanglement via a common bath is possible for bosonic and fermionic
baths.\cite{PhysRevLett.89.277901,solenov:035134} For spin baths
such a possibility has been explored for non-interacting spin
bath\cite{hamdouni:245323} as well as interacting
ones\cite{yuan:045331,yi:054102,lucamarini:062308,jing:174425} but
is restricted to uniform coupling models. It will be shown later in
this paper that induced entanglement is possible for local coupling
model considered in this work. We note that the induced entanglement
is also closely related to recent proposals of quantum communication
and teleportation via spin
chain.\cite{PhysRevLett.91.207901,venuti:060401}

Another important issue is the disentanglement dynamics of an
initially entangled state. It has attracted much attention in recent
years since Yu and Eberly\cite{yu:140404} and Jak\'obczyk and
Jamr\'oz\cite{Jakobczyk2004} predicted that two initially entangled
state without interaction can become completely disentangled at
finite time. This feature has been termed entanglement sudden death
(ESD). ESD has been studied theoretically within various models
\cite{santos:040305,al-qasimi:012117,ikram:062336} and has been
demonstrate experimentally.\cite{Almeida2007} These models, however,
are restricted to Markovian bosonic bath or classical noise. In this
work we explore for the first time if ESD-like phenomena can occur
for spin baths in non-Markovian regime. In particular, we will show
that only when the spin bath are paramagnetic or antiferromagnetic,
the phenomenon of ESD occurs, while when the spin bath is in the
ferromagnetic phase, concurrence shows oscillatory behavior.
As understanding the nature of the (dis)-entanglement dynamics constitutes an important
step for quantum engineering these systems, our results are of
practical usage for future quantum information processing.

The paper is organized as follows. In Sec. \ref{sec:model} we
present our model Hamiltonian and briefly discuss how to apply
t-DMRG to analyze the model Hamiltonian. In Sec. \ref{sec:single} we
present our results of single qubit decoherence while in Sec.
\ref{sec:two} the results of (dis)-entanglement dynamics are presented. In
Sec.\ref{sec:summary}, we summarize and discuss implication of our
results.

%%%%%%%%%%%%%%%%%%%%%%%%%%%%%%%%%%%%%%%%
\section{Theoretical Formulation}
\label{sec:model} We consider a system-bath model which is described
by the total Hamiltonian $H=H_{sys}+H_{bath}+H_{int}$, where
$H_{sys}$ is the Hamiltonian of a single or two qubits system,
$H_{bath}$ is the Hamiltonian of a spin bath and $H_{int}$
represents the interaction between qubits and the bath. We shall set
$H_{sys}=0$ but our method can be applied to a generic $H_{sys}$. We
shall assume that the spin bath is a spin chain characterized by the
$XXZ$ Heisenberg model
\begin{equation}
  \mathcal{H}_{bath} =
  J \sum \left( S^x_iS^x_{i+1}+S^y_iS^y_{i+1} +\Delta S^z_iS^z_{i+1} \right),
\end{equation}
where $J > 0$. It is known that the $XXZ$ Heisenberg model has a
very rich structure.\cite{Sachdev2000} The system is ferromagnetic
for $\Delta < 1$, antiferromagnetic (Ising-type) for $\Delta >1$,
and critical (XY-type) for $-1 < \Delta < 1$. It also encompasses
the $XY$ model where $\Delta=0$. The most general linear coupling
between qubit $A(B)$ and the bath can be expressed as
$H_{int}=\sum_{i,\alpha} \epsilon^{\alpha}_i  s_{A(B)}^\alpha
S^\alpha_i$, where $\alpha=x,y,z$ and $i=1,\dots,N$. Here
$\epsilon_i$ characterizes the coupling of the spin qubit to the
$i$th spin in the spin chain. For most situations, the more
interesting cases are $\epsilon < 0$
\cite{PhysRevA.66.052317,Yuan2007} and hence we shall concentrate on
negative $\epsilon$. Our numerical method, however, can be applied
equally well to cases with positive $\epsilon$. In our work, both
the Ising coupling and isotropic Heisenberg coupling will be
considered. The Ising coupling ($\epsilon^x_i=\epsilon^y_i=0$) gives
rise to a pure dephasing model while the isotropic Heisenberg
coupling ($\epsilon^x_i=\epsilon^y_i=\epsilon^z_i \neq 0$) induces
both dephasing and energy relaxation. The range of the coupling is
crucial for characterizing the interaction of qubits to the spin
bath. For uniform coupling $\epsilon_i$ is independent of $i$. This
is unrealistic but for uniform coupling the Loschmidt echo and the
entanglement dynamics can be calculated exactly by using
Jordan-Wigner transformation when the spin bath is the $XY$ model
($\Delta=0$).\cite{yi:054102,ou2007} However, the more realistic
coupling model is the local coupling model in which only the
coupling to the closest spin is nonvanishing. Nonetheless, there is
no analytic solutions known for this model. The reduced dynamics is
less studied but is more relevant to real experiments. In this case,
if the spin bath is ferromagnetic, the qubit couples directly to the
order parameter (the magnetization); while if the spin bath is
antiferromagnetic or paramagnetic, the qubit does not couple to the
order parameter. Hence the reduced dynamics exhibits completely
different behavior in different phases. As the local coupling model
is more relevant to real experiments, in the following we shall
concentrate on the local coupling model.

We now briefly outline the procedure to evaluate the reduced
dynamics of the qubits and other derived quantities. For a given set
of parameters, we first employ static DMRG\cite{white:10345} to find
the ground state $|G\rangle$ of the spin chain, where open boundary
condition (OBC) is used. We assume that at $t=0$ the initial total
state is a product state of the form:
$|\Phi(0)\rangle=|\psi_{sys}(0)\rangle|G\rangle$, where
$|\psi_{sys}(0)\rangle$ is some particular system state that we are
interested in. Formally the evolution of the reduced density matrix
can be obtained by evolving first the total state
\begin{equation}
  |\Phi(t)\rangle=e^{-iHt}|\psi_{sys}(0)\rangle|G\rangle,
\end{equation}
then tracing off the spin bath
\begin{equation}
  \rho_{sys}(t)=\mathrm{Tr}_{bath}|\Phi(t)\rangle\langle\Phi(t)|.
\end{equation}
Loschmidt echo and concurrence then can be evaluated from
$\rho_{sys}(t)$. In general evolving such a state is a formidable
task. t-DMRG, however, provides a way to efficiently evolve such a
state with high accuracy for a quasi-one dimensional system. We note
that the degrees of freedom of the qubits are kept exactly during
the t-DMRG calculation by targeting an appropriate state. The
dimension of the truncated Hilbert space is set to be $D=100$. For
short time decay simulation we set $J\delta t=10^{-3}$ in the
Trotter slicing while for entanglement dynamics we set $J\delta
t=0.1-0.5$ to balance the Trotter error and truncation error.

% In this work a minor modification to the standard td-DMRG procedure is introduced because
% we want to exactly keep track of the dynamics of the qubits.
% Using single qubit case as an example, we express the total state at time $t$ as
% \begin{equation}
%   |\Phi(t)\rangle
%   =a_+(t)|+\rangle|\Psi_+(t)\rangle
%   +a_-(t)|-\rangle|\Psi_-(t)\rangle,
% \end{equation}
% where $|\pm\rangle$ are the computational basis of the qubit.
% A second order Susuki-Trotter formula is then used as in the standard td-DMRG.
% At a typical step during the sweeping one has
% \begin{equation}
%   e^{-iH_i dt} \sum_\pm a_\pm|\Psi_\pm\rangle
%   = \sum_\pm a^\prime_\pm|\Psi^\prime_\pm\rangle
% \end{equation}
% where $H_i$ is the local Hamiltonian acting on bond-$i$.
% The new adaptive optimal DMRG basis is then obtained via targeting $|\Psi^\prime_\pm\rangle$
% with some appropriate weight (taken to be 1/2 and 1/2 in this work).

%%%%%%%%%%%%%%%%%%%%%%%%%%%%%%%%%%%%%%%%
\section{\label{sec:single}Single qubit decoherence}

In this section, we present our results for the single qubit
decoherence which is characterized by the Loschmidt echo. The
Loschmidt echo has been used extensively to quantify the single
qubit decoherence, especially its connection to the quantum
criticality of the spin baths.
\cite{quan:140604,cucchietti:032337,rossini:032333} Loschmidt echo
can be understood intuitively as follows: Consider an initially
disentangled total state $\left(C_+|+\rangle+C_-|-\rangle\right)
\otimes |G\rangle$, at some later time $t$ it will evolve into an
entangled state
$C_+(t)|+\rangle\otimes|\Psi_+(t)\rangle+C_-(t)|-\rangle\otimes|\Psi_-(t)\rangle$.
Loschmidt echo, defined as $\mathcal{L}(t)\equiv |\langle
\Psi_+(t)|\Psi_-(t)\rangle|^2$, clearly measure the coherence
between $|+\rangle$ and $|-\rangle$. When $\mathcal{L}=1$ the qubit
is disentangled from the bath while when $\mathcal{L}=0$ the qubit
is totally entangled with the bath.

\begin{figure}[htbp]
  \centering
  \includegraphics[width=3in]{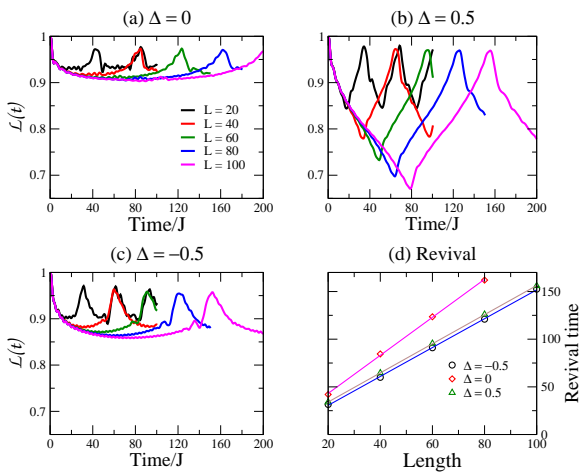}
  \caption{\label{fig:revival-i}(a)(b)(c) $\mathcal{L}(t)$ as a function of time for different lengths
    and $\Delta$. (d)Revival time as a function of length. The coupling is of Ising type with $\epsilon=-0.3$.}
\end{figure}

\begin{figure}[htbp]
  \centering
  \includegraphics[width=3in]{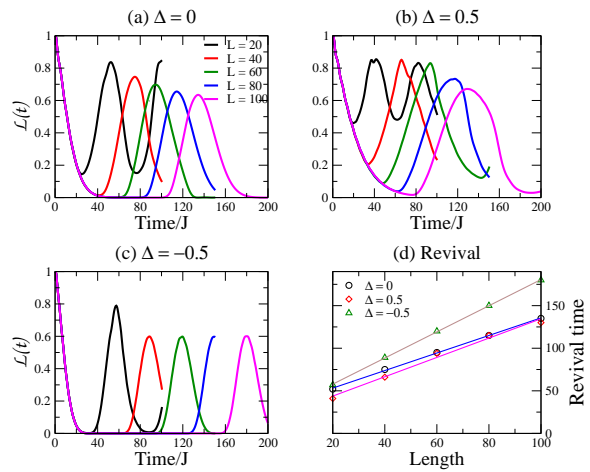}
  \caption{\label{fig:revival-h}(a)(b)(c) $\mathcal{L}(t)$ as a function of time for different lengths
    and $\Delta$. (d)Revival time as a function of length. The coupling is of Heisenberg type with $\epsilon=-0.3$.}
\end{figure}

We start by noting that for numerical calculation on finite length,
all dynamics will show quasi-periodic behavior. The quasi-period is
known as the revival time. Since in our numerical calculation, the
spin bath is a chain of finite length, it is essential to identify
the revival time for each length to avoid unphysical results due to
revival. As a zeroth order approximation, the revival time is
proportional to the length and inverse proportional to the maximum
phase velocity of the spin chain. In Figs.~\ref{fig:revival-i} and
\ref{fig:revival-h}, we plot the Loschmidt echo as a function of
time for the case of Ising and Heisenberg coupling using various
lengths and $\Delta$, from which the revival time of the echo can be
easily identified. We also plot the revival time as a function of
the length. One clearly observes the liner dependence of revival time on
the length. We find that for Ising coupling the minimal value
Loschmidt echo reached is non-zero unless a very strong coupling
strength is taken (not shown here), while for Heisenberg coupling,
the Loschmidt echo reaches zero if the length is longer than some
$\Delta$ dependent critical length.

To compute the single qubit decoherence, we couple the qubit to a
single site of spin chain, which is taken to be the middle site of
the chain to suppress boundary effects. We tune the spin bath to
different quantum phases by changing the parameter $\Delta$. In the
ferromagnetic regime, a small uniform external field is introduced
during the static-DMRG calculation but is turned off during time
evolution. It is numerically checked that the numerical results
reported below are insensitive to the magnitude of the applied
external field. Similarly, when the bath is in the Ising
antiferromagnetic ground state of the XXZ model, a small staggered
external field is applied to lift the two-fold degeneracy in the
ground state.\cite{Mikeska2004}

\begin{figure}
    \includegraphics[width=3in]{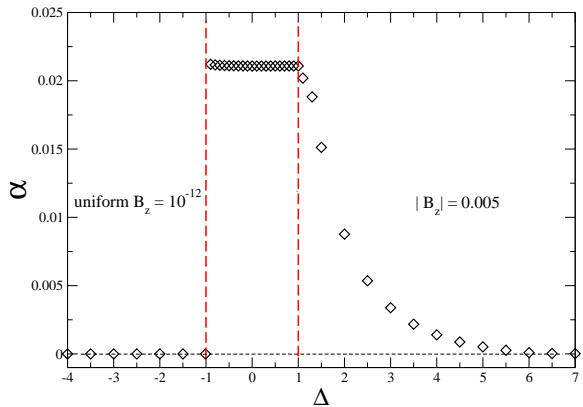}
  \caption{\label{fig:LE-Z}Decay parameter as a function of $\Delta$ for the case of Ising coupling. Here $\epsilon=-0.3$,
    and length $N=80$.}
\end{figure}

When the spin is in the Ising antiferromagnetic regime or the XY
critical regime, we find that in short time the behavior of the Loschmidt
echo decay is Gaussian, $\mathcal{L}(t) \sim e^{-\alpha t^2}$, where
$\alpha$ is the decay parameter.\cite{PhysRevA.30.1610}
%Known results in the literatures are, however, mostly restricted to pure dephasing models.
%In this work we not only report results for both pure dephasing and generic decoherence model
%but also provide data for more general coupling strength distributions.
In Fig.~\ref{fig:LE-Z} we plot the decay parameter as a function of
$\Delta$ for the case of Ising coupling.  In the ferromagnetic
regime ($\Delta < -1$), the qubit decay is completely suppressed
($\alpha=0$). This is a consequence of the Ising coupling in which
both  $ |+\rangle \otimes |G\rangle$ and $ |-\rangle \otimes
|G\rangle$ are eigenstates to the system and hence
$\mathcal{L}(t)=1$.

\begin{figure}
  \includegraphics[width=3in]{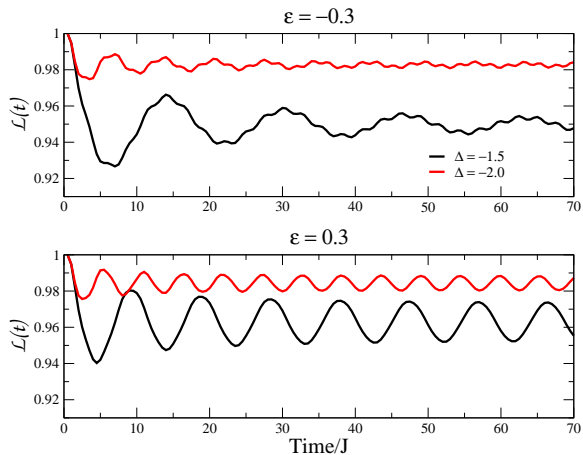}
  \caption{\label{fig:positive}$\mathcal{L}(t)$ as a function of time when the spin bath is in the ferromagnetic phase
    for the case of Heisenberg coupling. Here length $N=80$ and the coupling strength $\epsilon=-0.3$ (upper) or
    $\epsilon=0.3$ (lower) Clear oscillatory behaviors are seen for both positive and negative $\epsilon$.}
\end{figure}

\begin{figure}
  \includegraphics[width=3in]{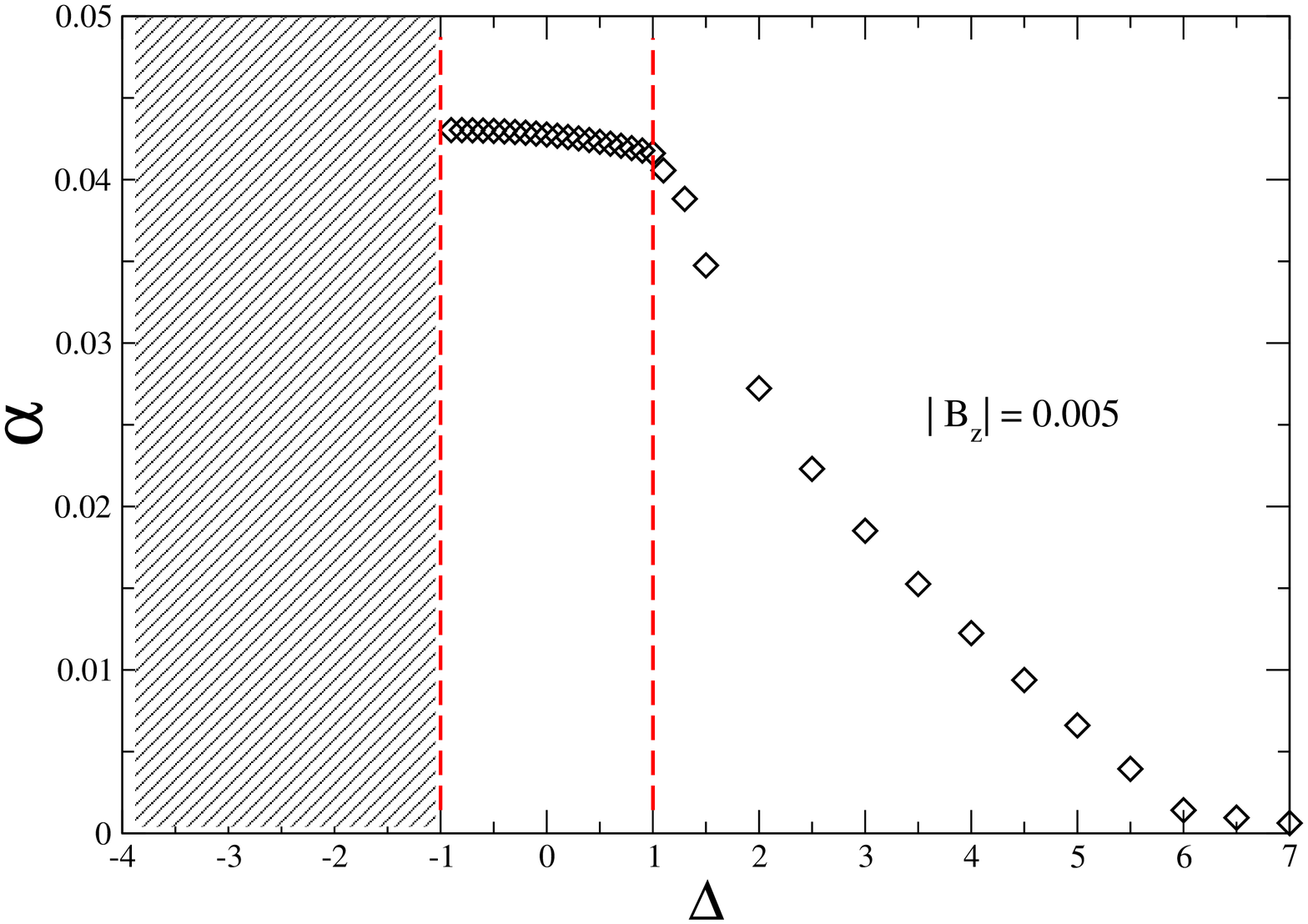}
  \caption{\label{fig:LE-XYZ}Decay parameter as a function of $\Delta$ for the case of Heisenberg coupling,
    $\epsilon=-0.3$, and length $N=80$}
\end{figure}
Clearly, the decay parameter is largest in the critical regime ($-1<
\Delta <1$) and it decreases gradually to zero as one moves into the
antiferromagnetic regime ($\Delta >1$). For single link scenario,
the decay parameter is almost featureless within the critical
regime. Our numerical results also show that if the qubit is coupled
to multiple sites, decay parameter acquires a weak dependence on
$\Delta$ and the transition near $\Delta=1$ becomes less sharp (not plotted).
%It is reported in the literature that in critical regime the decay parameter is a constant.\cite{rossini:032333}
%Our calculation shows that it is an artifact due to the single link assumption.
%We observe that for multiple link coupling the decay parameter is no longer a constant
%in the critical regime. The curve of decay parameter is discontinuous at $\Delta=-1$ and
%is continuous at $\Delta=1$ but with a discontinuity in its slope.
%When the width of Gaussian multi lik is increased, i.e.,  when the qubit is coupled to more sites,
%the change of slope becomes smaller and the decay be suppressed.
Note that the decay parameter becomes sensitive to the magnitude of
the small staggered field applied when the spin bath is close to the
phase boundary $(\Delta \sim 1)$. This is due to the fact that
for finite $N$ the barrier between two degenerate ground states is
finite and approaches zero as $\Delta$ approaches $1$. The ground
state obtained by static DMRG includes a small mixture of the
degenerate state, which is sensitive to the strength of the
staggered field. For larger $\Delta$, the barrier between two
degenerate ground states increases as one moves deep into the
antiferromagnetic regime. As a result, the decay parameter becomes
less sensitive to the strength of the stagger field for $\Delta \gg 1$.

We now turn to the case of Heisenberg coupling. We first note that
in the ferromagnetic regime, the qubit couples to the order
parameter. Therefore, the qubit is effectively in an average
magnetic field $\langle \vec{S}_i \rangle$. As a consequence of the
Heisenberg coupling, the qubit will precess about $\langle \vec{S}_i
\rangle$. Since magnons are generated at the same time when the
qubit evolves, $\langle \vec{S}_i \rangle$ starts to deviate from
$1/2$ and results in oscillations in the reduced dynamics.
Fig.~\ref{fig:positive} shows some typical oscillating behavior of
$\mathcal{L}(t)$ in this scenario. Clearly, the reduced dynamics is no longer
Gaussian. Therefore, we shall not mark the ferromagnetic regime in
the followings.

In Fig.~\ref{fig:LE-XYZ} we plot the decay parameter as a function
of $\Delta$ for the case of Heisenberg coupling. The overall
behavior is very similar to the case of Ising coupling except that
the decay parameter depends weakly on $\Delta$ in the critical
regime. This is different from the Ising coupling case shown above
but is similar to the multiple sites Ising coupling case. For both Ising
and Heisenberg coupling we find a discontinuity in the behavior of
$\mathcal{L}(t)$ at $\Delta=-1$ and a first derivative discontinuity
at $\Delta=1$. These discontinuities coincides with the phase
boundary of the underlying spin chain. Different behavior at the
$\Delta=\pm 1$ can be traced back to the different nature of the
ferromagnetic and antiferromagnetic transition and the close
relation between the decoherence and the quantum criticality of the
bath is clearly demonstrated.

%%%%%%%%%%%%%%%%%%%%%%%%%%%%%%%%%%%%%%%%
\section{\label{sec:two}Entanglement dynamics}
In this section, we investigate the entanglement dynamics of two
qubits that couple to the spin bath. There are two central issues to
be addressed. The first issue is the possibility of entanglement
creation via the common spin bath for a pair of initially
disentangled qubits without direct interaction. The second issue is
the disentanglement dynamics of an initially entangled state. In
particular, we would like to address the issue if qubits influenced
by spin baths also suffer from entanglement sudden death and if the
entanglement sudden death depends on the quantum phase that qubits
couple to.\cite{yu:140404}
%Would two state with the same initial entanglement undergo different disentanglement process,
%or even suffer from entanglement sudden death?\cite{yu:140404}
To characterize the entanglement, we shall use concurrence as the
measurement of entanglement.\cite{Wootters1998} For a given reduced
density matrix $\rho(t)$, the concurrence is defined as
$C=\max\{\lambda_1-\lambda_2-\lambda_3-\lambda_4,0\}$, where
$\lambda_1\ge\lambda_2\ge\lambda_3\ge\lambda_4$ are the square roots
of the eigenvalues of the operator
$\rho(\sigma^y\otimes\sigma^y)\rho^*(\sigma^y\otimes\sigma^y)$ and
$\rho^*$ is the complex conjugation of $\rho$.

\begin{figure}[htbp]
    \centering
        \includegraphics[width=3in]{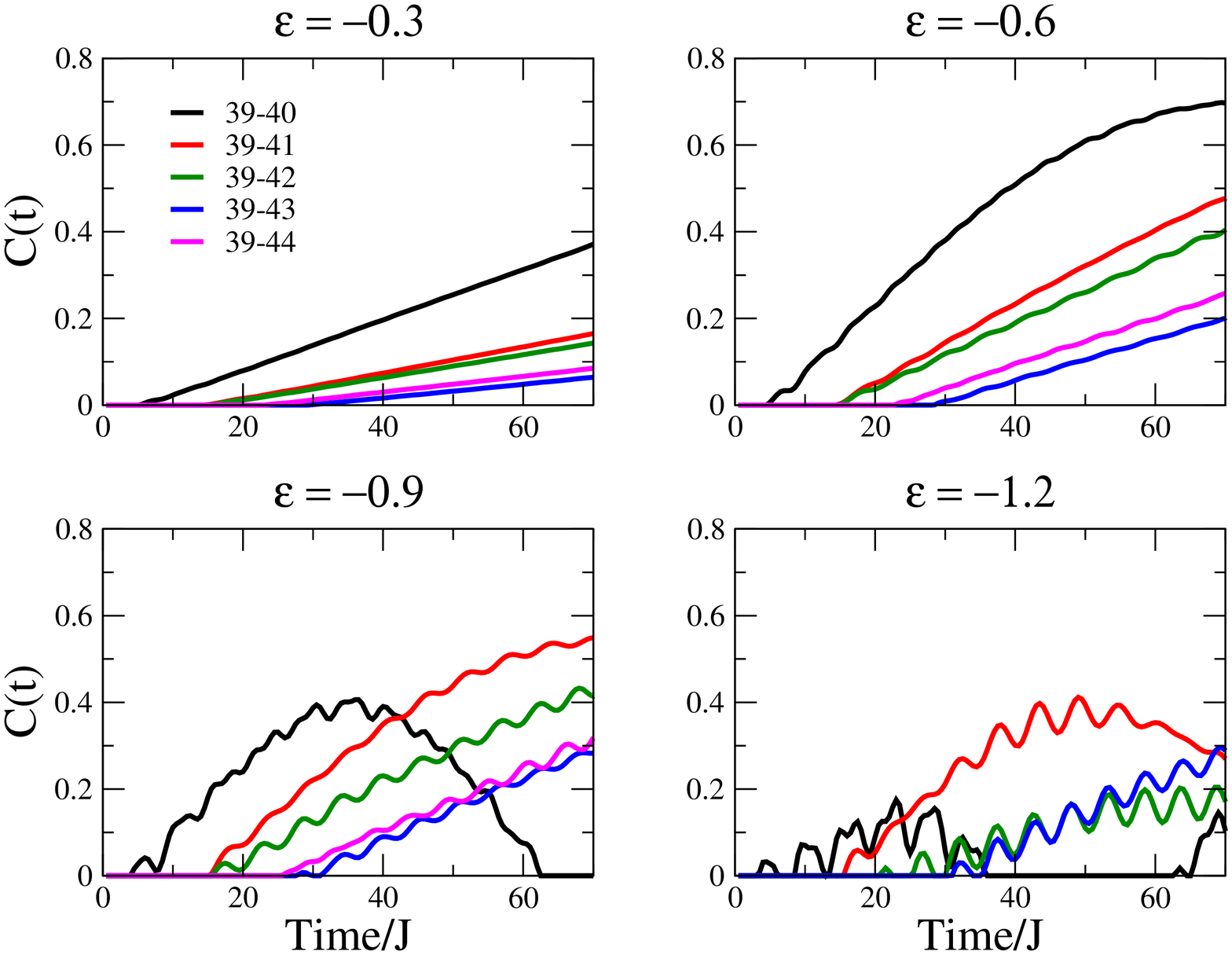}
    \caption{\label{fig:ent-c-i}Entanglement dynamics for an initially disentangled pair of qubits for the case
    of Ising coupling Here $\Delta=0$ and $N=80$.}
\end{figure}

\begin{figure}[htbp]
    \centering
        \includegraphics[width=3in]{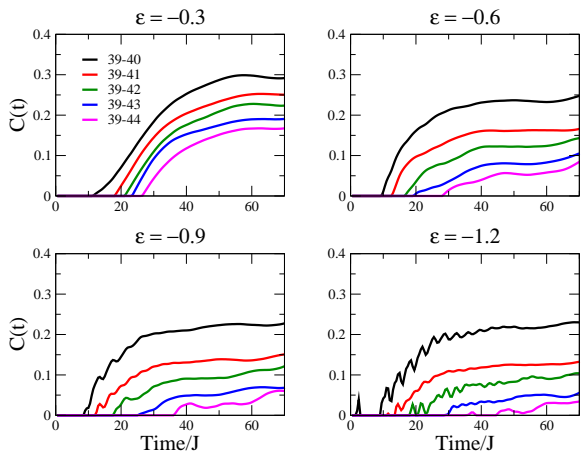}
        \caption{\label{fig:ent-c-h}Entanglement dynamics for an initially disentangled pair of qubits for the case
        of Heisenberg coupling. Here $\Delta=0$ and $N=80$.}
\end{figure}

\subsection{Entanglement creation}
It has been shown that entanglement can be created without direct
interaction if two qubits interact with a common bosonic bath
\cite{PhysRevLett.89.277901} or a fermionic
bath.\cite{solenov:035134} Of particular interest to us is the onset
time of the entanglement, the strength of the induced entanglement,
and the time scale where the induced decoherence eventually takes
over. These considerations are important in determining if such an
induced entanglement is useful in real quantum computation. The
issue is also closely related to the proposals of induced
interaction via a common bath,
\cite{PhysRevLett.89.167402,PhysRevLett.86.5112} where the effect of
induced decoherence from the same bath is usually neglected during
the derivation.

In Fig.~\ref{fig:ent-c-i}
we plot the concurrence as a function of time using various coupling strength and inter qubit distance.
The coupling between qubits and the spin bath is of Ising type, which gives rise to a pure dephasing model.
We assume that the coupling strength is the same for two qubits ($\epsilon_1=\epsilon_2$) and
the initial state is taken to be $\frac{1}{\sqrt{4}}(|00\rangle+|01\rangle+|10\rangle+|11\rangle)$.
We shall set $\Delta=0$ but similar results can ba obtained for $\Delta\neq 0$.
Before we discuss our findings in more details we would like to comment that
if Markovian approximation or uniform coupling assumption are taken then one can no longer
discuss the inter-qubit distance dependence. The relation between entanglement dynamics and inter-qubit distance,
however, is gaining interest since people begin to explore the non-Markovian effects of the bath.
\cite{chou:011112,solenov:035134,Anastopoulos2006}
We terminate the simulation at half of the revival time, where usually Loschmidt echo reaches
its minimum, to avoid the unphysical dynamics due to the revival.
We also numerically check the finite size effect by comparing the entanglement dynamics from different chain lengths.
We find that the results from different chain lengths agree with each other reasonably well. The length of the chain
mainly set an upper bound for the simulation time.

We find that for this configuration it is possible to create
entanglement via the spin bath. In particular, for weaker coupling
strength the induced entanglement rises more slowly but can reach a
higher value; while for stronger coupling the induced entanglement
rises more rapidly. The maximal concurrence reached, however, is
lower. This is because larger coupling strength also lead to
stronger decoherence. It is also evident from the figure that the
entanglement creating rate decreases as the inter qubit distance
increases, which is typical for this kind of induced interaction. We
find that for large enough $\epsilon$ and the smaller enough inter
qubit distance the concurrence shows oscillatory behavior. In these
cases, the coupling is strong enough to create concurrence
oscillation but also weak enough to prevent the bath to totally
disentangle the qubits. The delicate interplay between induced
decoherence and induced entanglement indicates that using such an
induced entanglement for quantum computation is a tricky task. One
has to tune the coupling to be within the right window to balance
the effect from each side.

In Fig.~\ref{fig:ent-c-h} we plot the concurrence as a function of
time for the case of Heisenberg coupling, starting form the same
initial condition. Qualitatively the behavior is similar to the case
of Ising coupling. We find that the maximal entanglement that can be
reached is smaller. This is because for Heisenberg coupling the
Loschmidt echo always decays to zero regardless the coupling
strength while for Ising coupling the minimal Loschmidt echo value
is a decreasing function of the coupling strength and is not zero.
We also find that the onset time is roughly proportional to the
inter qubit distance. This is expected as the excitation of spin
chain, which mediate the entanglement generation, travels with
finite phase velocity. The time the excitation reaches the other
qubit would be proportional to the inter qubit distance. However,
the concurrence oscillation is absent, indicating that the induced
interaction is weaker for Heisenberg coupling. We note that it is
difficult to write down an exact form of the induced interaction
unless Markovian approximation is taken. In general the induced
interaction is time-dependent and is accompanied by a complicate
decoherence effect. It is, however, possible to perform quantum
state tomography experimentally or numerically to extract Kraus
operators. The Kraus operators can then be used to design quantum
operations without using directly the form of the induced
interaction.

\begin{figure}
  \includegraphics[width=3in]{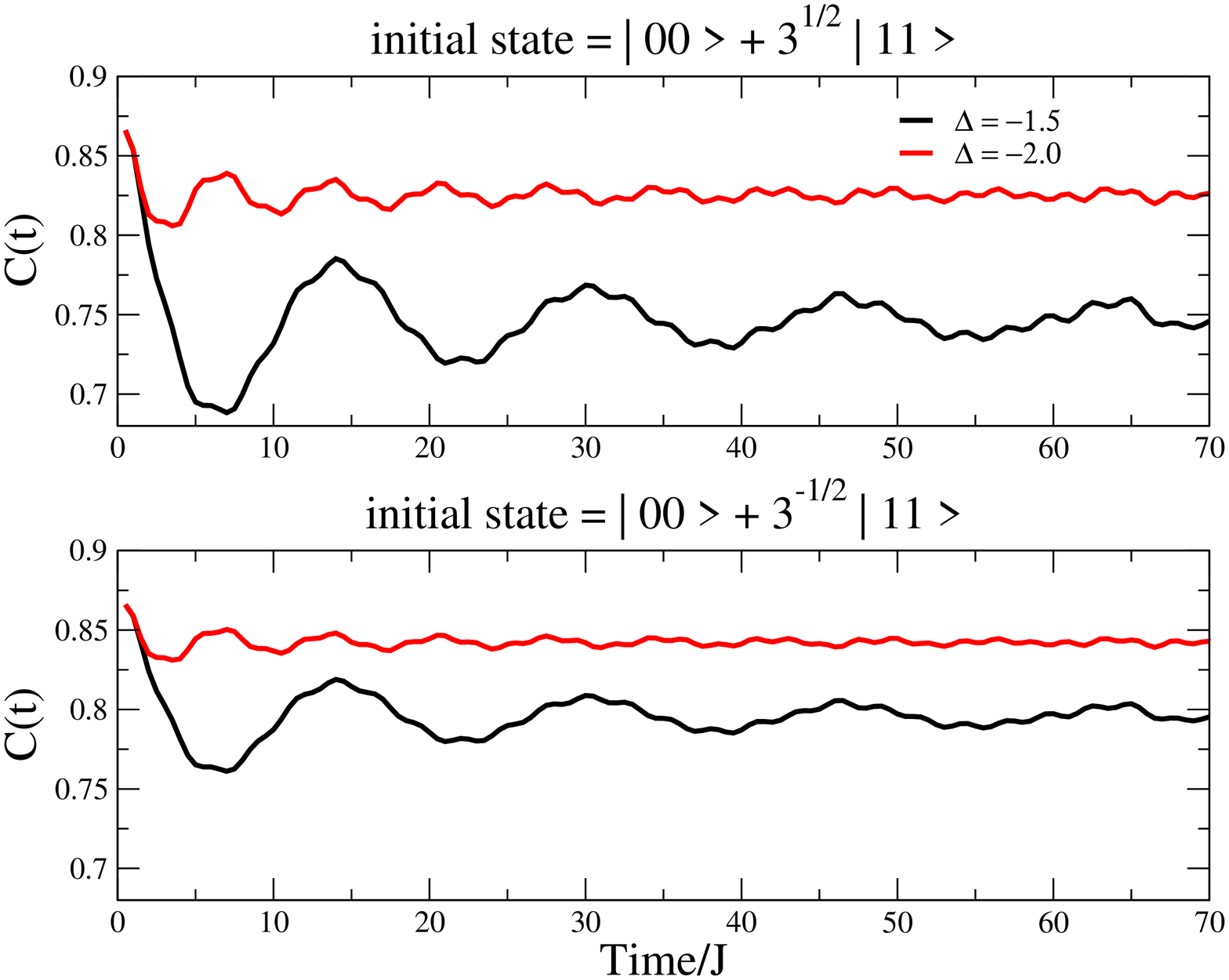}
  \caption{\label{fig:positivecc}Disentanglement dynamics for an initially disentangled pair of qubits
    when the bath is in the ferromagnetic phase. The coupling is of Heisenberg type, $\epsilon=-0.3$ and $N=80$.}
\end{figure}

\begin{figure}[htbp]
    \centering
        \includegraphics[width=3in]{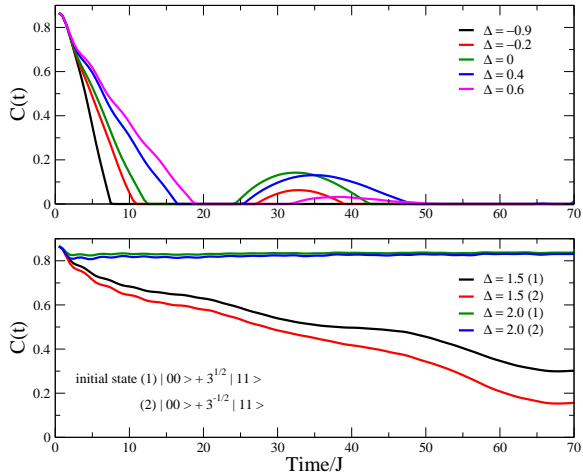}
    \caption{\label{fig:ESD}Disentanglement dynamics for an initially disentangled pair of qubits
                when the bath is in the antiferromagnetic or the XY critical phase.
                The coupling is of Heisenberg type, $\epsilon=-0.3$, and $N=80$.}
\end{figure}

\begin{figure}[htbp]
    \centering
        \includegraphics[width=3in]{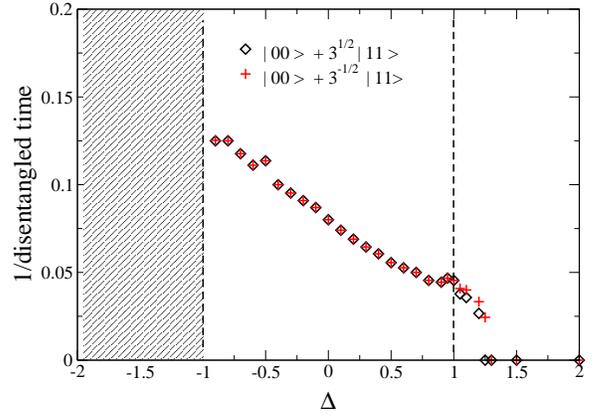}
    \caption{\label{fig:ESD-phase}Inverse finite disentanglement time as a function of $\Delta$.
        Here the coupling is the Heisenberg type, $\epsilon=-0.3$, and $N=80$.}
\end{figure}

\subsection{Entanglement decay}
Here we present our results for the disentanglement dynamics of an
initially entangled state. To investigate the possibility of ESD in
spin bath we start from an initial state of the form
$|\psi_{sys}(0)\rangle=\alpha|00\rangle+\beta|11\rangle$, with an
initial concurrence $C(0)=2|\alpha\beta^*|$. Two qubits are set be
20 sites apart so that the decoherence of individual qubit are
nearly independent and the coupling is of Heisenberg type.
We first show a typical behavior of concurrence in the ferromagnetic regime
in Fig.~\ref{fig:positivecc}. Clearly, as for
the reduced dynamics for a single qubit, the concurrence shows
oscillatory behaviors. Hence qubits in the ferromagnetic regime do
not suffer from ESD and the envelop of the entanglement decays exponentially.

In Fig.\ref{fig:ESD} we plot the disentanglement dynamics of two
states in the XY critical and antiferromagnetic regimes. All of the
computations start from the same initial concurrence with two set of
coefficients, $\alpha/\beta=1/\sqrt{3}$ or
$\beta/\alpha=1/\sqrt{3}$, corresponding to two different initial
states. We find that in the critical regime $(-1 < \Delta < +1)$
both states suffers from ESD. Furthermore, the entanglement dynamics
of two states are identical in the critical regime, which is due to
that the rotational symmetry is not broken in the critical regime
and $|00\rangle$ is equivalent to $|11\rangle$ by $Z_2$ symmetry
along the quantization axis. In both ferromagnetic and
antiferromagnetic phase, where the rotational symmetry is broken, we
find that the entanglement dynamics for these two states starts to
deviate from each other. In most of the antiferromagnetic regime
both states do not suffer from ESD. When $\Delta$ is close to the
phase boundary, however, both of two states that correspond to
$\alpha/\beta=1/\sqrt{3}$ and $\beta/\alpha=1/\sqrt{3}$ suffer from
ESD and have slightly different disentanglement time. In
Fig.~\ref{fig:ESD-phase} we plot the inverse of disentanglement
time, which is defined as the time when concurrence becomes zero, as
a function of $\Delta$. Starting from $\Delta=-1$, it shows
monotonic decrease. Across the phase boundary, $\Delta=1$, the
inverse of disentanglement time develops a small bump and persists
into the antiferromagentic phase and finally decreases to zero
around $\Delta=1.2$. The existence of finite region with finite
disentanglement time in the antiferromagnetic regime is due to that
when $\Delta$ approaches $1$, the barrier between two degenerate
ground states approaches zero. For finite $N$ and finite value of
$\Delta -1$, the ground state obtained by static DMRG includes a
small mixture of the degenerate state so that it resembles the XY
critical state and results in finite disentanglement time. In the
thermodynamic limit ($N \rightarrow \infty$), the region with finite
disentanglement time in the antiferromagnetic regime shrinks down,
resulting discontinuity at the phase boundary $\Delta=1$. The
overall behavior found in the above is different from those reported
in the early work\cite{santos:040305} in which two states
investigated in the above posses different entanglement dynamics and
only one of them suffer from ESD. The difference is due to that the
model adopted in Ref.~\cite{santos:040305} includes the effect of
spontaneous decay, breaking the symmetry between $|0\rangle$ and
$|1\rangle$; while for spin bath in XY critical regime such symmetry
breaking is absent. It is important to note that there exits a
subregime, roughly around $-0.3 < \Delta < 0.6$, in which
entanglement shows revival after some dark period. Note that the
entanglement revival after some dark period was also reported in
Ref.~\cite{ficek:024304} where a photonic multimode vacuum bath is
assumed and the revival is attributed to the two photon decay. In
our work, the origin of the revival is less clear. We believe that
the existence of such sub-regime is due to the competition between
entanglement decay and Loschmidt echo decay. Within this subregime
the Loschmidt echo decay more slowly, giving the system a chance to
revive after the first ESD.

% \begin{figure}[htbp]
%     \centering
%         \includegraphics[width=3in]{rest.eps}
%     \caption{ESD}
% \end{figure}
% \begin{figure}[htbp]
%   \centering
%       \includegraphics[width=3in]{fig10.eps}
%   \caption{ESD}
% \end{figure}

%%%%%%%%%%%%%%%%%%%%%%%%%%%%%%%%%%%%%%%%
\section{\label{sec:summary} Conclusion}

In summary, the decoherence and (dis)-entanglement dynamics induced
by spin baths are investigated non-perturbatively by using t-DMRG.
For both pure dephasing model (Ising coupling) and general
decoherence model (Heisenberg coupling) we calculate the short time
decay parameter of the Loschmidt echo. We find that in both cases
the decay parameter is closely related to the phase of the underlying
spin chain. In the ferromagnetic regime, the reduced dynamics shows
oscillatory behavior while in the XY critical and
antiferromagnetic regimes, the decay parameter shows a first
derivative discontinuity at $\Delta=+1$. We evaluate the
entanglement dynamics of a pair of initially disentangled qubits
which are close to each other. We demonstrate that it is
possible to induced entanglement via their common interaction with
the spin bath. The competition between induced decoherence and
entanglement can be easily seen in the coupling strength dependent
behavior of the entanglement onset time, growth rate, and the
maximal entanglement reached. Finally we investigate the
disentanglement dynamics of a pair initially entangled qubits which
are far from each other. For the two initial states we studied, we
find that their disentanglement dynamics are identical and suffer
from ESD in the critical regime. Their disentanglement dynamics
begin to deviate from each other in both the ferromagnetic and the
antiferromagnetic regime. They no longer suffer from ESD in the
ferromagnetic regime but still suffer from ESD if the chain is near
the antiferromagnetic transition. It is shown that the inverse of
finite disentanglement time has a close relation to the phase of the
spin bath and shows monotonic decrease behavior as one moves into
the antiferromagnetic regime.

We thank Prof. Hsiu-Hau Lin for inspiring discussions and comments.
This work is supported by National Science Council of Taiwan.

% \begin{figure}[htbp]
%     \centering
%         \includegraphics[width=3in]{tlh_stagger.eps}
%     \caption{\label{fig:xxx}L(t) for case of stagger coupling in AFM}
% \end{figure}
%
% \begin{figure}[htbp]
%     \centering
%         \includegraphics[width=3in]{ttlh_entanglement_fm.eps}
%     \caption{\label{fig:yyy}C(t) in AFM, local coupling}
% \end{figure}

%\bibliography{../../phys.bib}

\begin{thebibliography}{42}
\expandafter\ifx\csname natexlab\endcsname\relax\def\natexlab#1{#1}\fi
\expandafter\ifx\csname bibnamefont\endcsname\relax
  \def\bibnamefont#1{#1}\fi
\expandafter\ifx\csname bibfnamefont\endcsname\relax
  \def\bibfnamefont#1{#1}\fi
\expandafter\ifx\csname citenamefont\endcsname\relax
  \def\citenamefont#1{#1}\fi
\expandafter\ifx\csname url\endcsname\relax
  \def\url#1{\texttt{#1}}\fi
\expandafter\ifx\csname urlprefix\endcsname\relax\def\urlprefix{URL }\fi
\providecommand{\bibinfo}[2]{#2}
\providecommand{\eprint}[2][]{\url{#2}}

\bibitem[{\citenamefont{Khaetskii and Nazarov}(2000)}]{PhysRevB.61.12639}
\bibinfo{author}{\bibfnamefont{A.~V.} \bibnamefont{Khaetskii}}
  \bibnamefont{and} \bibinfo{author}{\bibfnamefont{Y.~V.}
  \bibnamefont{Nazarov}}, \bibinfo{journal}{Phys. Rev. B}
  \textbf{\bibinfo{volume}{61}}, \bibinfo{pages}{12639} (\bibinfo{year}{2000}).

\bibitem[{\citenamefont{Ladd et~al.}(2005)\citenamefont{Ladd, Maryenko,
  Yamamoto, Abe, and Itoh}}]{ladd:014401}
\bibinfo{author}{\bibfnamefont{T.~D.} \bibnamefont{Ladd}},
  \bibinfo{author}{\bibfnamefont{D.}~\bibnamefont{Maryenko}},
  \bibinfo{author}{\bibfnamefont{Y.}~\bibnamefont{Yamamoto}},
  \bibinfo{author}{\bibfnamefont{E.}~\bibnamefont{Abe}}, \bibnamefont{and}
  \bibinfo{author}{\bibfnamefont{K.~M.} \bibnamefont{Itoh}},
  \bibinfo{journal}{Phys. Rev. B} \textbf{\bibinfo{volume}{71}},
  \bibinfo{eid}{014401} (\bibinfo{year}{2005}).

\bibitem[{\citenamefont{Loss and DiVincenzo}(1998)}]{PhysRevA.57.120}
\bibinfo{author}{\bibfnamefont{D.}~\bibnamefont{Loss}} \bibnamefont{and}
  \bibinfo{author}{\bibfnamefont{D.~P.} \bibnamefont{DiVincenzo}},
  \bibinfo{journal}{Phys. Rev. A} \textbf{\bibinfo{volume}{57}},
  \bibinfo{pages}{120} (\bibinfo{year}{1998}).

\bibitem[{\citenamefont{Stamp and Tupitsyn}(2004)}]{Stamp2004}
\bibinfo{author}{\bibfnamefont{P.~C.~E.} \bibnamefont{Stamp}} \bibnamefont{and}
  \bibinfo{author}{\bibfnamefont{I.~S.} \bibnamefont{Tupitsyn}},
  \bibinfo{journal}{Chem. Phys.} \textbf{\bibinfo{volume}{296}},
  \bibinfo{pages}{281} (\bibinfo{year}{2004}).

\bibitem[{\citenamefont{Tian et~al.}(2007)\citenamefont{Tian, Chen, and
  Wang}}]{arXiv:0707.1131}
\bibinfo{author}{\bibfnamefont{Y.-Y.} \bibnamefont{Tian}},
  \bibinfo{author}{\bibfnamefont{P.}~\bibnamefont{Chen}}, \bibnamefont{and}
  \bibinfo{author}{\bibfnamefont{D.-W.} \bibnamefont{Wang}},
  \bibinfo{journal}{arxiv}  (\bibinfo{year}{2007}), \eprint{arXiv-0707.1131}.

\bibitem[{\citenamefont{Cucchietti et~al.}(2005)\citenamefont{Cucchietti, Paz,
  and Zurek}}]{cucchietti:052113}
\bibinfo{author}{\bibfnamefont{F.~M.} \bibnamefont{Cucchietti}},
  \bibinfo{author}{\bibfnamefont{J.~P.} \bibnamefont{Paz}}, \bibnamefont{and}
  \bibinfo{author}{\bibfnamefont{W.~H.} \bibnamefont{Zurek}},
  \bibinfo{journal}{Phys. Rev. A} \textbf{\bibinfo{volume}{72}},
  \bibinfo{eid}{052113} (\bibinfo{year}{2005}).

\bibitem[{\citenamefont{Breuer et~al.}(2004)\citenamefont{Breuer, Burgarth, and
  Petruccione}}]{breuer:045323}
\bibinfo{author}{\bibfnamefont{H.-P.} \bibnamefont{Breuer}},
  \bibinfo{author}{\bibfnamefont{D.}~\bibnamefont{Burgarth}}, \bibnamefont{and}
  \bibinfo{author}{\bibfnamefont{F.}~\bibnamefont{Petruccione}},
  \bibinfo{journal}{Phys. Rev. B} \textbf{\bibinfo{volume}{70}},
  \bibinfo{eid}{045323} (\bibinfo{year}{2004}).

\bibitem[{\citenamefont{Yuan et~al.}(2007{\natexlab{a}})\citenamefont{Yuan,
  Goan, and Zhu}}]{yuan:045331}
\bibinfo{author}{\bibfnamefont{X.-Z.} \bibnamefont{Yuan}},
  \bibinfo{author}{\bibfnamefont{H.-S.} \bibnamefont{Goan}}, \bibnamefont{and}
  \bibinfo{author}{\bibfnamefont{K.-D.} \bibnamefont{Zhu}},
  \bibinfo{journal}{Phys. Rev. B} \textbf{\bibinfo{volume}{75}},
  \bibinfo{eid}{045331} (\bibinfo{year}{2007}{\natexlab{a}}).

\bibitem[{\citenamefont{Paganelli et~al.}(2002)\citenamefont{Paganelli,
  de~Pasquale, and Giampaolo}}]{PhysRevA.66.052317}
\bibinfo{author}{\bibfnamefont{S.}~\bibnamefont{Paganelli}},
  \bibinfo{author}{\bibfnamefont{F.}~\bibnamefont{de~Pasquale}},
  \bibnamefont{and} \bibinfo{author}{\bibfnamefont{S.~M.}
  \bibnamefont{Giampaolo}}, \bibinfo{journal}{Phys. Rev. A}
  \textbf{\bibinfo{volume}{66}}, \bibinfo{pages}{052317}
  (\bibinfo{year}{2002}).

\bibitem[{\citenamefont{Yuan et~al.}(2007{\natexlab{b}})\citenamefont{Yuan,
  Goan, and Zhu}}]{Yuan2007}
\bibinfo{author}{\bibfnamefont{X.-Z.} \bibnamefont{Yuan}},
  \bibinfo{author}{\bibfnamefont{H.-S.} \bibnamefont{Goan}}, \bibnamefont{and}
  \bibinfo{author}{\bibfnamefont{K.-D.} \bibnamefont{Zhu}},
  \bibinfo{journal}{New Journal of Physics} \textbf{\bibinfo{volume}{9}},
  \bibinfo{pages}{219} (\bibinfo{year}{2007}{\natexlab{b}}), ISSN
  \bibinfo{issn}{1367-2630}.

\bibitem[{\citenamefont{Vidal}(2004)}]{vidal:040502}
\bibinfo{author}{\bibfnamefont{G.}~\bibnamefont{Vidal}},
  \bibinfo{journal}{Phys. Rev. Lett.} \textbf{\bibinfo{volume}{93}},
  \bibinfo{eid}{040502} (\bibinfo{year}{2004}).

\bibitem[{\citenamefont{White and Feiguin}(2004)}]{white:076401}
\bibinfo{author}{\bibfnamefont{S.~R.} \bibnamefont{White}} \bibnamefont{and}
  \bibinfo{author}{\bibfnamefont{A.~E.} \bibnamefont{Feiguin}},
  \bibinfo{journal}{Phys. Rev. Lett.} \textbf{\bibinfo{volume}{93}},
  \bibinfo{eid}{076401} (\bibinfo{year}{2004}).

\bibitem[{\citenamefont{Rossini et~al.}(2007)\citenamefont{Rossini, Calarco,
  Giovannetti, Montangero, and Fazio}}]{rossini:032333}
\bibinfo{author}{\bibfnamefont{D.}~\bibnamefont{Rossini}},
  \bibinfo{author}{\bibfnamefont{T.}~\bibnamefont{Calarco}},
  \bibinfo{author}{\bibfnamefont{V.}~\bibnamefont{Giovannetti}},
  \bibinfo{author}{\bibfnamefont{S.}~\bibnamefont{Montangero}},
  \bibnamefont{and} \bibinfo{author}{\bibfnamefont{R.}~\bibnamefont{Fazio}},
  \bibinfo{journal}{Phys. Rev. A} \textbf{\bibinfo{volume}{75}},
  \bibinfo{eid}{032333} (\bibinfo{year}{2007}).

\bibitem[{\citenamefont{Gobert et~al.}(2005)\citenamefont{Gobert, Kollath,
  Schollwock, and Schutz}}]{gobert:036102}
\bibinfo{author}{\bibfnamefont{D.}~\bibnamefont{Gobert}},
  \bibinfo{author}{\bibfnamefont{C.}~\bibnamefont{Kollath}},
  \bibinfo{author}{\bibfnamefont{U.}~\bibnamefont{Schollwock}},
  \bibnamefont{and} \bibinfo{author}{\bibfnamefont{G.}~\bibnamefont{Schutz}},
  \bibinfo{journal}{Phys. Rev. E} \textbf{\bibinfo{volume}{71}},
  \bibinfo{eid}{036102} (\bibinfo{year}{2005}).

\bibitem[{\citenamefont{Cucchietti et~al.}(2003)\citenamefont{Cucchietti,
  Dalvit, Paz, and Zurek}}]{PhysRevLett.91.210403}
\bibinfo{author}{\bibfnamefont{F.~M.} \bibnamefont{Cucchietti}},
  \bibinfo{author}{\bibfnamefont{D.~A.~R.} \bibnamefont{Dalvit}},
  \bibinfo{author}{\bibfnamefont{J.~P.} \bibnamefont{Paz}}, \bibnamefont{and}
  \bibinfo{author}{\bibfnamefont{W.~H.} \bibnamefont{Zurek}},
  \bibinfo{journal}{Phys. Rev. Lett.} \textbf{\bibinfo{volume}{91}},
  \bibinfo{pages}{210403} (\bibinfo{year}{2003}).

\bibitem[{\citenamefont{Wootters}(1998)}]{Wootters1998}
\bibinfo{author}{\bibfnamefont{W.~K.} \bibnamefont{Wootters}},
  \bibinfo{journal}{Phys. Rev. Lett.} \textbf{\bibinfo{volume}{80}},
  \bibinfo{pages}{2245} (\bibinfo{year}{1998}).

\bibitem[{\citenamefont{Braun}(2002)}]{PhysRevLett.89.277901}
\bibinfo{author}{\bibfnamefont{D.}~\bibnamefont{Braun}},
  \bibinfo{journal}{Phys. Rev. Lett.} \textbf{\bibinfo{volume}{89}},
  \bibinfo{pages}{277901} (\bibinfo{year}{2002}).

\bibitem[{\citenamefont{Solenov et~al.}(2007)\citenamefont{Solenov, Tolkunov,
  and Privman}}]{solenov:035134}
\bibinfo{author}{\bibfnamefont{D.}~\bibnamefont{Solenov}},
  \bibinfo{author}{\bibfnamefont{D.}~\bibnamefont{Tolkunov}}, \bibnamefont{and}
  \bibinfo{author}{\bibfnamefont{V.}~\bibnamefont{Privman}},
  \bibinfo{journal}{Phys. Rev. B} \textbf{\bibinfo{volume}{75}},
  \bibinfo{eid}{035134} (\bibinfo{year}{2007}).

\bibitem[{\citenamefont{Hamdouni et~al.}(2006)\citenamefont{Hamdouni, Fannes,
  and Petruccione}}]{hamdouni:245323}
\bibinfo{author}{\bibfnamefont{Y.}~\bibnamefont{Hamdouni}},
  \bibinfo{author}{\bibfnamefont{M.}~\bibnamefont{Fannes}}, \bibnamefont{and}
  \bibinfo{author}{\bibfnamefont{F.}~\bibnamefont{Petruccione}},
  \bibinfo{journal}{Phys. Rev. B} \textbf{\bibinfo{volume}{73}},
  \bibinfo{eid}{245323} (\bibinfo{year}{2006}).

\bibitem[{\citenamefont{Yi et~al.}(2006)\citenamefont{Yi, Cui, and
  Wang}}]{yi:054102}
\bibinfo{author}{\bibfnamefont{X.~X.} \bibnamefont{Yi}},
  \bibinfo{author}{\bibfnamefont{H.~T.} \bibnamefont{Cui}}, \bibnamefont{and}
  \bibinfo{author}{\bibfnamefont{L.~C.} \bibnamefont{Wang}},
  \bibinfo{journal}{Phys. Rev. A} \textbf{\bibinfo{volume}{74}},
  \bibinfo{eid}{054102} (\bibinfo{year}{2006}).

\bibitem[{\citenamefont{Lucamarini et~al.}(2004)\citenamefont{Lucamarini,
  Paganelli, and Mancini}}]{lucamarini:062308}
\bibinfo{author}{\bibfnamefont{M.}~\bibnamefont{Lucamarini}},
  \bibinfo{author}{\bibfnamefont{S.}~\bibnamefont{Paganelli}},
  \bibnamefont{and} \bibinfo{author}{\bibfnamefont{S.}~\bibnamefont{Mancini}},
  \bibinfo{journal}{Phys. Rev. A} \textbf{\bibinfo{volume}{69}},
  \bibinfo{eid}{062308} (\bibinfo{year}{2004}).

\bibitem[{\citenamefont{Jing and Lu}(2007)}]{jing:174425}
\bibinfo{author}{\bibfnamefont{J.}~\bibnamefont{Jing}} \bibnamefont{and}
  \bibinfo{author}{\bibfnamefont{Z.-G.} \bibnamefont{Lu}},
  \bibinfo{journal}{Phys. Rev. B} \textbf{\bibinfo{volume}{75}},
  \bibinfo{eid}{174425} (\bibinfo{year}{2007}).

\bibitem[{\citenamefont{Bose}(2003)}]{PhysRevLett.91.207901}
\bibinfo{author}{\bibfnamefont{S.}~\bibnamefont{Bose}}, \bibinfo{journal}{Phys.
  Rev. Lett.} \textbf{\bibinfo{volume}{91}}, \bibinfo{pages}{207901}
  (\bibinfo{year}{2003}).

\bibitem[{\citenamefont{Venuti et~al.}(2007)\citenamefont{Venuti, Boschi, and
  Roncaglia}}]{venuti:060401}
\bibinfo{author}{\bibfnamefont{L.~C.} \bibnamefont{Venuti}},
  \bibinfo{author}{\bibfnamefont{C.~D.~E.} \bibnamefont{Boschi}},
  \bibnamefont{and}
  \bibinfo{author}{\bibfnamefont{M.}~\bibnamefont{Roncaglia}},
  \bibinfo{journal}{Phys. Rev. Lett.} \textbf{\bibinfo{volume}{99}},
  \bibinfo{eid}{060401} (\bibinfo{year}{2007}).

\bibitem[{\citenamefont{Yu and Eberly}(2004)}]{yu:140404}
\bibinfo{author}{\bibfnamefont{T.}~\bibnamefont{Yu}} \bibnamefont{and}
  \bibinfo{author}{\bibfnamefont{J.~H.} \bibnamefont{Eberly}},
  \bibinfo{journal}{Phys. Rev. Lett.} \textbf{\bibinfo{volume}{93}},
  \bibinfo{eid}{140404} (\bibinfo{year}{2004}).

\bibitem[{\citenamefont{Jakobczyk and Jamroz}(2004)}]{Jakobczyk2004}
\bibinfo{author}{\bibfnamefont{L.}~\bibnamefont{Jakobczyk}} \bibnamefont{and}
  \bibinfo{author}{\bibfnamefont{A.}~\bibnamefont{Jamroz}},
  \bibinfo{journal}{Physics Letters A} \textbf{\bibinfo{volume}{333}},
  \bibinfo{pages}{35} (\bibinfo{year}{2004}).

\bibitem[{\citenamefont{Santos et~al.}(2006)\citenamefont{Santos, Milman,
  Davidovich, and Zagury}}]{santos:040305}
\bibinfo{author}{\bibfnamefont{M.~F.} \bibnamefont{Santos}},
  \bibinfo{author}{\bibfnamefont{P.}~\bibnamefont{Milman}},
  \bibinfo{author}{\bibfnamefont{L.}~\bibnamefont{Davidovich}},
  \bibnamefont{and} \bibinfo{author}{\bibfnamefont{N.}~\bibnamefont{Zagury}},
  \bibinfo{journal}{Phys. Rev. A} \textbf{\bibinfo{volume}{73}},
  \bibinfo{eid}{040305} (\bibinfo{year}{2006}).

\bibitem[{\citenamefont{Al-Qasimi and James}(2008)}]{al-qasimi:012117}
\bibinfo{author}{\bibfnamefont{A.}~\bibnamefont{Al-Qasimi}} \bibnamefont{and}
  \bibinfo{author}{\bibfnamefont{D.~F.~V.} \bibnamefont{James}},
  \bibinfo{journal}{Phys. Rev. A} \textbf{\bibinfo{volume}{77}},
  \bibinfo{eid}{012117} (\bibinfo{year}{2008}).

\bibitem[{\citenamefont{Ikram et~al.}(2007)\citenamefont{Ikram, li~Li, and
  Zubairy}}]{ikram:062336}
\bibinfo{author}{\bibfnamefont{M.}~\bibnamefont{Ikram}},
  \bibinfo{author}{\bibfnamefont{F.}~\bibnamefont{li~Li}}, \bibnamefont{and}
  \bibinfo{author}{\bibfnamefont{M.~S.} \bibnamefont{Zubairy}},
  \bibinfo{journal}{Phys. Rev. A} \textbf{\bibinfo{volume}{75}},
  \bibinfo{eid}{062336} (\bibinfo{year}{2007}).

\bibitem[{\citenamefont{Almeida et~al.}(2007)\citenamefont{Almeida, de~Melo,
  Hor-Meyll, Salles, Walborn, Ribeiro, and Davidovich}}]{Almeida2007}
\bibinfo{author}{\bibfnamefont{M.~P.} \bibnamefont{Almeida}},
  \bibinfo{author}{\bibfnamefont{F.}~\bibnamefont{de~Melo}},
  \bibinfo{author}{\bibfnamefont{M.}~\bibnamefont{Hor-Meyll}},
  \bibinfo{author}{\bibfnamefont{A.}~\bibnamefont{Salles}},
  \bibinfo{author}{\bibfnamefont{S.~P.} \bibnamefont{Walborn}},
  \bibinfo{author}{\bibfnamefont{P.~H.~S.} \bibnamefont{Ribeiro}},
  \bibnamefont{and}
  \bibinfo{author}{\bibfnamefont{L.}~\bibnamefont{Davidovich}},
  \bibinfo{journal}{Science} \textbf{\bibinfo{volume}{316}},
  \bibinfo{pages}{579} (\bibinfo{year}{2007}).

\bibitem[{\citenamefont{Sachdev}(2000)}]{Sachdev2000}
\bibinfo{author}{\bibfnamefont{S.}~\bibnamefont{Sachdev}},
  \emph{\bibinfo{title}{Quantum Phase Transition}}
  (\bibinfo{publisher}{Cambridge University Press, Cambridge},
  \bibinfo{year}{2000}).

\bibitem[{\citenamefont{Ou and Fan}(2007)}]{ou2007}
\bibinfo{author}{\bibfnamefont{Y.-C.} \bibnamefont{Ou}} \bibnamefont{and}
  \bibinfo{author}{\bibfnamefont{H.}~\bibnamefont{Fan}}, \bibinfo{journal}{J.
  Phys. A: Math. Theor.} \textbf{\bibinfo{volume}{40}}, \bibinfo{pages}{2455}
  (\bibinfo{year}{2007}).

\bibitem[{\citenamefont{White}(1993)}]{white:10345}
\bibinfo{author}{\bibfnamefont{S.~R.} \bibnamefont{White}},
  \bibinfo{journal}{Phys. Rev. B} \textbf{\bibinfo{volume}{48}},
  \bibinfo{pages}{10345} (\bibinfo{year}{1993}).

\bibitem[{\citenamefont{Quan et~al.}(2006)\citenamefont{Quan, Song, Liu,
  Zanardi, and Sun}}]{quan:140604}
\bibinfo{author}{\bibfnamefont{H.~T.} \bibnamefont{Quan}},
  \bibinfo{author}{\bibfnamefont{Z.}~\bibnamefont{Song}},
  \bibinfo{author}{\bibfnamefont{X.~F.} \bibnamefont{Liu}},
  \bibinfo{author}{\bibfnamefont{P.}~\bibnamefont{Zanardi}}, \bibnamefont{and}
  \bibinfo{author}{\bibfnamefont{C.~P.} \bibnamefont{Sun}},
  \bibinfo{journal}{Phys. Rev. Lett.} \textbf{\bibinfo{volume}{96}},
  \bibinfo{eid}{140604} (\bibinfo{year}{2006}).

\bibitem[{\citenamefont{Cucchietti et~al.}(2007)\citenamefont{Cucchietti,
  Fernandez-Vidal, and Paz}}]{cucchietti:032337}
\bibinfo{author}{\bibfnamefont{F.~M.} \bibnamefont{Cucchietti}},
  \bibinfo{author}{\bibfnamefont{S.}~\bibnamefont{Fernandez-Vidal}},
  \bibnamefont{and} \bibinfo{author}{\bibfnamefont{J.~P.} \bibnamefont{Paz}},
  \bibinfo{journal}{Phys. Rev. A} \textbf{\bibinfo{volume}{75}},
  \bibinfo{eid}{032337} (\bibinfo{year}{2007}).

\bibitem[{\citenamefont{Peres}(1984)}]{PhysRevA.30.1610}
\bibinfo{author}{\bibfnamefont{A.}~\bibnamefont{Peres}},
  \bibinfo{journal}{Phys. Rev. A} \textbf{\bibinfo{volume}{30}},
  \bibinfo{pages}{1610} (\bibinfo{year}{1984}).

\bibitem[{\citenamefont{Mikeska and Kolezhuk}(2004)}]{Mikeska2004}
\bibinfo{author}{\bibfnamefont{H.-J.} \bibnamefont{Mikeska}} \bibnamefont{and}
  \bibinfo{author}{\bibfnamefont{A.~K.} \bibnamefont{Kolezhuk}},
  \emph{\bibinfo{title}{Quantum Magnetism}}, vol. \bibinfo{volume}{645} of
  \emph{\bibinfo{series}{Lecture Notes in Physics}}
  (\bibinfo{publisher}{Springer Berlin / Heidelberg}, \bibinfo{year}{2004}).

\bibitem[{\citenamefont{Piermarocchi et~al.}(2002)\citenamefont{Piermarocchi,
  Chen, Sham, and Steel}}]{PhysRevLett.89.167402}
\bibinfo{author}{\bibfnamefont{C.}~\bibnamefont{Piermarocchi}},
  \bibinfo{author}{\bibfnamefont{P.}~\bibnamefont{Chen}},
  \bibinfo{author}{\bibfnamefont{L.~J.} \bibnamefont{Sham}}, \bibnamefont{and}
  \bibinfo{author}{\bibfnamefont{D.~G.} \bibnamefont{Steel}},
  \bibinfo{journal}{Phys. Rev. Lett.} \textbf{\bibinfo{volume}{89}},
  \bibinfo{pages}{167402} (\bibinfo{year}{2002}).

\bibitem[{\citenamefont{Mozyrsky et~al.}(2001)\citenamefont{Mozyrsky, Privman,
  and Glasser}}]{PhysRevLett.86.5112}
\bibinfo{author}{\bibfnamefont{D.}~\bibnamefont{Mozyrsky}},
  \bibinfo{author}{\bibfnamefont{V.}~\bibnamefont{Privman}}, \bibnamefont{and}
  \bibinfo{author}{\bibfnamefont{M.~L.} \bibnamefont{Glasser}},
  \bibinfo{journal}{Phys. Rev. Lett.} \textbf{\bibinfo{volume}{86}},
  \bibinfo{pages}{5112} (\bibinfo{year}{2001}).

\bibitem[{\citenamefont{Chou et~al.}(2008)\citenamefont{Chou, Yu, and
  Hu}}]{chou:011112}
\bibinfo{author}{\bibfnamefont{C.-H.} \bibnamefont{Chou}},
  \bibinfo{author}{\bibfnamefont{T.}~\bibnamefont{Yu}}, \bibnamefont{and}
  \bibinfo{author}{\bibfnamefont{B.~L.} \bibnamefont{Hu}},
  \bibinfo{journal}{Phys. Rev. E} \textbf{\bibinfo{volume}{77}},
  \bibinfo{eid}{011112} (\bibinfo{year}{2008}).

\bibitem[{\citenamefont{Anastopoulos et~al.}(2006)\citenamefont{Anastopoulos,
  Shresta, and Hu}}]{Anastopoulos2006}
\bibinfo{author}{\bibfnamefont{C.}~\bibnamefont{Anastopoulos}},
  \bibinfo{author}{\bibfnamefont{S.}~\bibnamefont{Shresta}}, \bibnamefont{and}
  \bibinfo{author}{\bibfnamefont{B.~L.} \bibnamefont{Hu}}
  (\bibinfo{year}{2006}), \eprint{arXiv:quant-ph/0610007}.

\bibitem[{\citenamefont{Ficek and Tana\'{s}}(2006)}]{ficek:024304}
\bibinfo{author}{\bibfnamefont{Z.}~\bibnamefont{Ficek}} \bibnamefont{and}
  \bibinfo{author}{\bibfnamefont{R.}~\bibnamefont{Tana\'{s}}},
  \bibinfo{journal}{Phys. Rev. A} \textbf{\bibinfo{volume}{74}},
  \bibinfo{eid}{024304} (\bibinfo{year}{2006}).

\end{thebibliography}

\end{document}